# Is symmetry identity?

Marvin Chester

Physics emeritus, UCLA, Los Angeles, California, USA

**Abstract**  *Wigner found unreasonable the "effectiveness of mathematics in the natural sciences". But if the mathematics we use to describe nature is simply a coded expression of our experience then its effectiveness is quite reasonable. Its effectiveness is built into its design. We consider group theory, the logic of symmetry. We examine the premise that symmetry is identity; that group theory encodes our experience of identification. To decide whether group theory describes the world in such an elemental way we catalogue the detailed correspondence between elements of the physical world and elements of the formalism. Providing an unequivocal match between concept and mathematical statement completes the case. It makes effectiveness appear reasonable. The case that symmetry is identity is a strong one but it is not complete. The further validation required suggests that unexpected entities might be describable by the irreducible representations of group theory.*

## 1. Effectiveness Speaks

In his famous paper entitled, "The Unreasonable Effectiveness of Mathematics in the Natural Sciences", Eugene Wigner (1979, p. 223) wrote:

> "... the enormous usefulness of mathematics in the natural sciences is something bordering on the mysterious ... there is no rational explanation for it."

Some think there might be a rational explanation. In his book *The Applicability of Mathematics as a Philosophical Problem,* Steiner (1998, p. 5) finds that the use of mathematics "cannot avoid being an anthropocentric strategy". It rests ultimately on the human experience of nature. He is thus led to "explore the implication for our view of the universe" (p. 2) of the evident applicability of mathematics to the physical world. Steiner turns Wigner's plaint inside out asking what the evident effectiveness of mathematics tells us. In what follows I will support and expand on this seminal notion.

Wigner was an acknowledged master of group theory, the mathematical theory of symmetry. Laws of nature - physical laws - are governed by group theory. Bas van Fraassen demonstrates this in his book, *Laws and Symmetry* (van Fraassen, 1989). He shows us that the status of 'physical law' is conferred by symmetry - invariance under the transformations of nature.

But what is symmetry that it should underlay the very foundation of natural law? Alternatively put: Why is group theory so effective in describing the physical world? The answer is that it codifies the basic axioms of the scientific enterprise. The logic of group theory is the logic of scientific inquiry so that the mathematics we use to describe nature is a carefully coded expression of our experience.

Group theory is the mathematical formulation of internal consistency in the description of things. We assume that the system being observed has an intrinsic character independent of the observer's perspective. It's there. It possesses an objective reality. On this assumption

- that it's there - how the system is perceived under altered scrutiny must be a matter of logic. Its appearance follows the logic of intrinsic sameness (Section 7). The codification of that logic is a matter of group theory. And its success in portraying the physical world is what validates the assumption.

Rather than ponder its efficacy we take instruction from the mathematics. We know - by experiment - that the physicist's mathematical description of the physical world is approximately correct. From the very efficacy of mathematics in describing it we may derive a message about the nature of the physical world. It is this inversion of Wigner's quest that we pursue here.

## 2. How symmetry is identity

I propose that, as used to describe the physical world, symmetry is so elemental that it coincides with the concept of identity itself. The theory of symmetry is the mathematical expression of the notion of identification and that is why it is so effective as the basis of science. By *identity* is meant the end result of *identification*, not the other sense of the word pertaining to identical.

That symmetry has played a substantive role in thinking about nature has a venerable history dating back to antiquity. This history is nicely outlined by Roche (1987). He discusses many examples. All of them demonstrate how symmetry considerations have helped solve physical problems. For example Descartes deduces that "two equal elastic bodies which collide with equal velocities rebound with these velocities exactly reversed because of symmetry". (Quote from Roche, 1987, p. 18)

It was Ernst Cassirer who first articulated the idea that the mathematical theory of symmetry - group theory - may transcend its problem-solving utility. Group theory "has not only a mathematical or physical but a universal epistemological interest .." he wrote (Cassirer, 1945, p. 273). The suggestion was that it has something to do with how we know; how we evaluate perceptions.

We know objects by their properties. Constitution is what "confers to the carrier of a set of properties the dignity of an object", says Elena Castellani (1998, p. 182). She 'constitutes' an elementary object of physics - electron, nucleon - from group theory by showing that invariance under the spatio-temporal transformation group yields as characterization of the object its energy, its linear momentum, its angular momentum and its mass. To constitute something, then, is to assign to it *labels of significance.* The significance arises from invariance properties.

This exercise exemplifies a broader view of symmetry. Although not explicitly stated the idea is implied - as it is in the work of others (Van Fraassen, 1989 and Toraldo di Francia, 1998)- that symmetry may be literally equated with identity. To constitute is to identify. We explore the notion that symmetry *is* identity; that group theory is the theory of identity.

If, indeed, group theory describes the world in such an elemental way then we must be able to provide the detailed correspondence between elements of the physical world and elements of the formalism. This is the way we accept Wigner's challenge to make reasonable the effectiveness of the mathematics. To display associations between mathematical notions and physical ones we need concise verbal expressions to match the concise mathematical ones. Expressions which generate images are the ones that make the mathematics *reasonable*.

Using non-technical language, here is a synopsis of the case. The essential quality that characterizes symmetry is this: the appearance of sameness under altered scrutiny. That this definition corresponds to group theory as applied to the physical world is grounded in Section 8. But the same phrase - a perceived sameness under altered scrutiny - is just what captures the notion of identity. When something is recognizably the same under many perspectives we grant it identity. An identification is made by labelling. The label tags what it is that we perceive as invariant. "If there were no invariants we could not define 'identity'" noted M.L.G. Redhead (1975, p. 78). Here we posit that it is precisely invariance that identifies identity.

Group Theory has an innate taxonomic structure - a taxonomy for behaviour. It assigns labels for behaviour (appearance) under altered scrutiny. The irreducible representation labels of Group Theory are the identification markers - the labels of identity. This is explored in Sections 9, 11 and 12. A central concept in this exposition is *altered scrutiny*. It is examined in Section 4. In applications to the physical world it is precisely *altered scrutinies* that *are the group elements.*

To ground the case synopsized in the preceding paragraphs we use a notation that iconizes the philosophical content: Dirac notation. It is the natural tool for the group theory of physical processes. What follows is a review of material that every group theory scholar knows, but recast so as to expose how the mathematics encodes our axioms about how nature works. To do this we focus on the correspondence between the physical world and the formalism.

## 3. The observer, the system, its states and its rules

In examining anything one is an observer. So in asking questions about nature we focus on what *an observer* might find. The observer divides the world into distinguishable parts. Each one, perceived as a discernibly distinct entity, may be called a 'system'. The system is what is studied by the observer; something accessible to measurement. The difficulty about the concept lies in the matter of isolation; that the system not be coupled to the rest of the universe. Evidently such systems are not to be found in nature. It is an idealization. Coupling is a matter of degree; never equal to zero. We follow tradition on the matter assuming coupling weak enough to qualify as zero. [1]

An isolated system is always to be found in a 'state'. A state is specified by its appearance by which is meant the results of a compatible set of measurements on the system. According to quantum mechanics these are eigenvalues of each of a complete set of commuting operators. [2] This is discussed further in Section 9. What is relevant here is that by 'appearance' is meant a set of numbers marking compatible measurement results. We label that set of numbers by a single integer, $n$. A state of the system is designated by $|n\rangle$, using standard Dirac notation.[3]

A system is governed by a rule. Its behaviour is determined by a law - or rule - through which its nature is expressed. In physics this takes the form of a Hamiltonian a Lagrangian or a variational principle.

The observer makes the measurements. He or she is equipped with a battery of instruments. The observer's eyes are one such instrument but he or she will usually have others - like clocks, meter sticks, electrometers, particle detectors, etc. In accordance with von Neumann's principle of Psychophysical Parallelism, it doesn't make any difference whether the instrument is within the observer's body or not. (See von Neumann, 1955, p. 419-420) He or she scrutinizes the system by recording the measurement results his instruments read. Making sure that all his measurements are compatible he or she assigns to a particular collection of his or her measurement results a particular value of $n$. He or she concludes that the system is in the state $|n\rangle$.

## 4. Altered scrutiny

The thread by which symmetry and identity are bound together is altered scrutiny. An altered scrutiny of the system means that the observer moves to another frame of reference to record his measurements. To this new reference frame he takes with him his entire ensemble of instrumentation - as if they were, indeed, a part of the observer's body. By altered scrutiny is meant the action that puts the observer into position to make his new measurements. It excludes the act of measurement itself - the scrutiny. The latter can change the state of the system. The former cannot.

The formal mathematical term for altered scrutiny is transformation. If what one means by *transformation* is *altered scrutiny* then the connection between identity and symmetry becomes clear. It is the observer's altered scrutiny that transforms the state. Altered scrutiny is purely an observer's construct. It is a reweaving of the fabric of descriptive space, an enterprise that goes on in the mind of the observer.

The prototypical change in reference frame is a physical rotation of the observer's coordinates. Figure 1 schematizes the idea. There the system's state is portrayed as a function in some descriptive space in which a representative vector is $\rho$. To each $\rho$ corresponds an amplitude $\langle \rho | n \rangle$ which measures the strength of the state at the point $\rho$. In the figure the function $\langle \rho | n \rangle$ is the pyramid whose projection we see.

The key feature of altered scrutiny is this: that corresponding to the observer's action of altering her scrutiny there is an operator $G_g$ whose effect is to produce a new state, $G_g | n \rangle$, from the state $| n \rangle$, observed originally. The new state, $G_g | n \rangle$, is called the transformed state. And $G_g$ is the transformation operator. This is illustrated in Figure 1. What $G_g$ produces when it acts on $| n \rangle$, we have called a 'state'. This proclaims a notion about reality: that altering one's perspective on a system cannot change its intrinsic nature. We see the same system, but in some other state. Its appearance is altered.

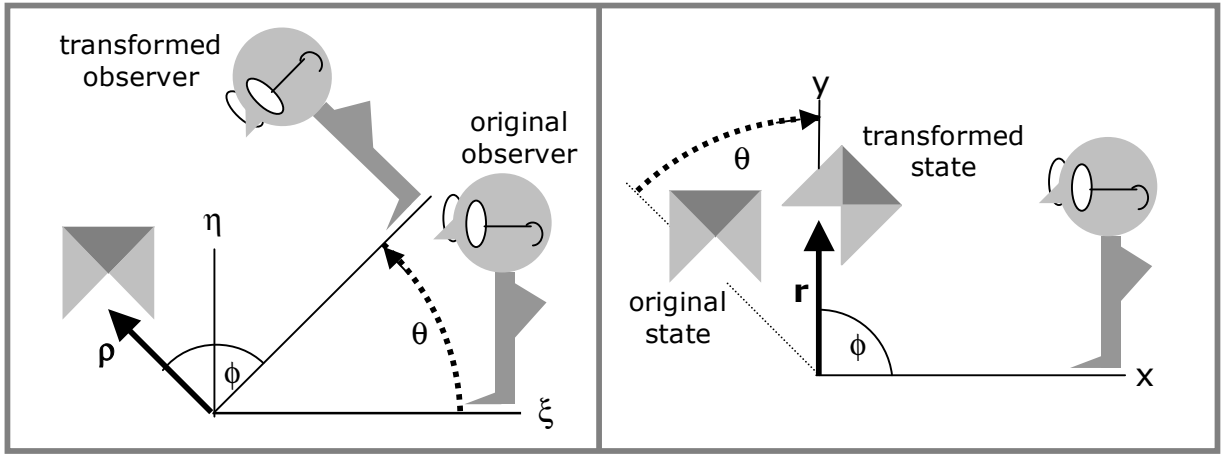

Figure 1. The observer's altered scrutiny is what transforms the state.

Whatever represents a physical system has the technical property that its states constitute an invariant[4] vector space for the group of all altered scrutinies. This is the formal way of saying that from a state of the system an altered scrutiny can only produce another state of the same system. We don't alter the system by looking at it. That is the significance of calling $G_g|n\rangle$ a 'state'. Mathematics allows other possibilities.[5] Not all vector spaces are invariant to group elements. So our notion about reality is expressed as a mathematical constraint. Its consequences are pursued in equation (4) of Section 10.

We enumerate altered scrutinies by subscripts. The g[th] altered scrutiny is $G_g$. The figure shows one of the altered scrutinies belonging to the continuous Lie group, SO(2). In that case we can use θ for g. The altered scrutiny displayed is $G_g = G_\theta$ or $G(\theta)$.

That altered scrutinies fall into groups follows from the definition. They conform to the five defining properties of a group. 1. Their law of combination is simply sequential action. Two alterations in scrutiny performed one after the other is written in product form, $G_b G_a$ where the order means $G_a$ is done first. 2. Closure is guaranteed because two alterations in scrutiny performed sequentially amounts itself to an altered scrutiny. The observer reorients himself and then, without taking measurements, does so again. He could just as well do it in one step. 3. Associativity obtains simply because there is no meaning to inserting parentheses among a product of altered scrutinies. They are simply a series of sequential actions. 4. All inverses exist. Simply undo the alteration in scrutiny. And 5. The identity element is the 'do nothing' action. Don't alter scrutiny. Thus altered scrutinies can always be grouped.

Altered scrutiny or transformation is something that produces a mapping of the states of the system onto themselves. In this generality it is not easily conceivable how an observer might physically execute many altered scrutinies. Inversion is an example. It may not be physically executable but it is conceptually executable: replace position variables by minus themselves. The observer alters his description of the system. That is his altered scrutiny. The phrase generates an image in the mind: those transformations that an observer can physically execute. We wish, in the definition, to embrace abstract altered scrutinies; transformations that one may not be able to execute physically but that one can imagine - that one can execute mathematically - including, say, changes in motion and even of phase in a wave function.

## 5. The system remains inviolate

Altered scrutiny actions have nothing to do with the system. They have only to do with the observer. The observer uses the same measuring instruments to make his measurements, but from a different perspective. In texts on group theory the distinction is made between the passive view and the active one.[6] They differ only in labeling. Aside from this they are mathematically equivalent. They are not equivalent philosophically.

What is traditionally called the passive view is shown at the left of Figure 1. The observer moves. The system is left inviolate. The active view is described as altering the physical system under inspection.[7] That is how one might interpret the right side of Figure 1. It is an interpretation we deliberately avoid. The right side of Figure 1 results - not from altering the system - but from altering the observer. The system's appearance is altered - but its integrity remains untouched. It is transformed but always left undisturbed.

The equivalence of the views mathematically assures us that in the use of group theory to describe a physical system we may always ascribe the transformation to a change in the observer's perspective. An example is the particle exchange transformation. It may be entirely reinterpreted as a relabeling of particles by the observer. (Fonda and Ghirardi, 1970, p.39) In this view the exchange is not of particles but of the enumeration labels used by the observer to tag them.

## 6. Descriptive space

When the observer alters his perspective he is readjusting his basis in descriptive space. In the figure his new basis is (x,y), his old is (ξ,η). The altered scrutiny action, $G_g$, is what produces the new basis from the old. If the physical point at position $\rho$ is described under altered scrutiny as being at **r** then the meaning of $G_g$ is embodied in the expression **r** $= G_g \rho$. By $G_g$ is meant, here, the prescription for delivering the appropriate **r** for the given $\rho$. In that way $G_g$ implements the altered scrutiny. This prescription, for the case in the figure, is best delivered in matrix form.

$$\begin{pmatrix} x \\ y \end{pmatrix} = G(\theta) \begin{pmatrix} \xi \\ \eta \end{pmatrix} = \begin{pmatrix} \cos\theta & \sin\theta \\ -\sin\theta & \cos\theta \end{pmatrix} \begin{pmatrix} \xi \\ \eta \end{pmatrix} \qquad (1)$$

This is a compact rendering of x=x(ξ,η) and y=y(ξ,η), the equations represented by

$\mathbf{r} = G_g \boldsymbol{\rho}$.

An observer can reverse his scrutiny alteration thus retrieving ρ from **r**. Put symbolically $\boldsymbol{\rho} = G_g^{-1}\mathbf{r}$. So a state of the system may equally well be portrayed in the transformed coordinates, **r**, by just carrying out the coordinate transformation; substituting $G_g^{-1}\mathbf{r}$ for ρ. The mathematical rendering of this statement is

$$\langle \boldsymbol{\rho} | n \rangle = \langle G_g^{-1}\mathbf{r} | n \rangle \qquad (2)$$

This notates the idea that altered scrutiny is a matter of rewriting the descriptive space basis in terms of the new basis coordinates.

## 7. The principle of intrinsic sameness

We now have two different prescriptions for the operator, $G_g$, that represents an alteration of an observer's perspective. It produces a transformed state, $G_g |n\rangle$, from the original $|n\rangle$ in Section 4. It generates a new descriptive space, **r**, from the old one, $G_g^{-1}\mathbf{r}$ in Section 6. That the two prescriptions be commensurate represents a philosophical commitment: that the altered appearance of a system (its transformed state) is due _only_ to the altered scrutiny of the observer (the transformed coordinate system) and to nothing else. That is the content of Equation (3).

$$\left\langle G_g^{-1}\mathbf{r}\,\middle|\,n\right\rangle = \langle \mathbf{r}|G_g|n\rangle \tag{3}$$

The left hand side gives the state's amplitude at a certain point of descriptive space. The right hand side insures that the transformed state has the same amplitude at the same physical point in that space. It ordains that the transformed state $G_g|n\rangle$ be what arises from altered perspective. It does this by giving us the prescription for discovering it computationally. Equation (3) is just that prescription. It is the mathematical formulation of our elemental intuition that a system has properties independent of our scrutiny; that even though the system appears different under altered scrutiny its intrinsic sameness is preserved. Thus, equation (3) expresses the logic of intrinsic sameness; that the system is <u>there</u> regardless of our scrutiny of it. The application of group theory to the physical world - to physics - is based on this equation. In texts it goes by the formal name of "induced transformation".[8] But this term fails to capture its formidable philosophical significance.

## 8. Symmetry is apparent sameness under altered scrutiny

We are now prepared to examine what it is that we mean by *symmetry*. What is the essential test by which we decide operationally that something is symmetric? Because the word is in everyone's vocabulary, a meaning in 'laymen's terms' often dissolves into a plethora of examples. And a definition becomes a matter of abstract mathematics.[9] We wish to define it in laymen's terms, succinctly but with a view to accommodating the mathematics because it is through simplicity and suggestive imagery that the mathematics appears reasonable.

Symmetry is *apparent sameness under altered scrutiny*. This phrase certainly captures the experience of our visual sense of symmetry. Visually any object that we call symmetrical has this key property: that under *some* altered scrutiny it looks the same to an observer. It is congruent to itself. In most cases, appearance changes when we alter our scrutiny. Among all the altered scrutinies possible only a few produce same appearance. Those few define the symmetry. But 'same appearance' need not be restricted to the visual. 'Same measurement results' qualify also. The definition is an operational one; scrutinize to check for sameness.

It's important to distinguish between *intrinsic* sameness and *apparent* sameness. Intrinsic sameness is always preserved. What constitutes symmetry is that an observer's altered scrutiny leave the object's *appearance unchanged*. When apparent sameness is preserved under some altered scrutiny, symmetry is present. The technical expression, 'invariant under the transformation' is what 'perceived sameness under altered scrutiny' means.

If there are some altered scrutinies then there are a group of them. We generate the group by applying already discovered same-appearance-scrutinies in sequence until all the members of the group emerge. Altered scrutinies are the group elements of group theory. These same-appearance-scrutinies refer, of course, to what are formally called 'covering operations' or 'symmetry operations'. Each same-appearance-scrutiny is the inverse of a symmetry operation. They differ in point of view: one moves the system, the other the observer.

## 9. Measurements provide enumeration labels

The states of a physical system are characterized by constellations of measurement results: the ones enumerated by n. As archetype we think of the spectrum of deflections (the states) of an atom beam (the system) passing through a magnetic field gradient as in a Stern-Gerlach experiment. (Chester, 1987, p. 148-150)  Physical states correspond to rays in Hilbert space.  Each state, $|n\rangle$, is a basis vector in this space. Thus Hilbert space is constructed from measurement results. What the observer sees - via his scrutiny apparatus - are a set of basis states in the Hilbert space of the system.[10]  Any state in the space is a superposition of the basis states.

## 10. Altered scrutinies generate group representations

What is perceived under altered scrutiny, $G_g |m\rangle$, is some state of the system. It is within the system's Hilbert space. Since any state of the system is a superposition of basis states so must be the altered scrutiny state. Put mathematically the statement is:

$$\langle \mathbf{r} | G_g | m \rangle = \sum_n \langle \mathbf{r} | n \rangle \langle n | G_g | m \rangle \qquad m = \text{any of the } n \qquad (4)$$

Formally this equation says that the Hilbert space with basis, $|n\rangle$, is invariant with respect to the group $\{G_g\}$. Physically it expresses the reasonable expectation that from a state of the system an altered scrutiny can produce only another state of the system. Because the observer carries his descriptive space with him, any state is describable in terms of his basis states, $\langle \mathbf{r} | n \rangle$, even the transformed state, $\langle \mathbf{r} | G_g | m \rangle$, that he encounters because of his alteration in scrutiny.

The coefficients in the sum - written $\langle n | G_g | m \rangle$ - represent the extent to which the transformed state $G_g |m\rangle$ is like $|n\rangle$. They are deducible computationally by incorporating the intrinsic sameness rule of equation (3) into equation (4). Choosing to write these coefficients as $\langle n|G_g|m\rangle$, exposes their nature: elements in a square matrix. The matrix is called a representation of the transformation. It is a mathematical rendering of the particular altered scrutiny action, $G_g$. When gathered for a group of such altered scrutinies the matrices of coefficients constitute a representation of that group. That is because multiplication of the corresponding matrices implements the sequential performance of altered scrutinies.

But these coefficients are connected to the statistics gathered in observations. They are experimentally accessible. The relative number of times an altered scrutiny of the state $|m\rangle$ will unearth the state $|n\rangle$ is just $|\langle n|G_g|m\rangle|^2$. So the group representation matrix elements, $\langle n|G_g|m\rangle$, are associated with measurement results.

An observer examining the states of a system cannot help but generate a matrix representation of the group as she goes through her group of altered scrutinies. The matrices in the representation will always be square and invertible. Invertible because an altered scrutiny always has an inverse. Square because both n and m belong to the same basis of states. A fundamental notion in group theory is that its mathematics is completely described in terms of invertible square matrices. Just such matrices arise naturally from examining the physical world.

## 11. Irreducible representations yield labels

The important thing about a matrix representation of a group is that it can be reduced to irreducible representations - to a direct sum of special symmetry matrices. These embody all the invariance properties of the group of transformations. Irreducible representations, $\Gamma_g^\gamma$, are matrices with elements, $\left\langle \kappa \left| \Gamma_g^\gamma \right| k \right\rangle$, where the subscript $g$, as earlier in $G_g$, marks a specific altered scrutiny. And $\gamma$ is the irreducible representation label - the *irrep label*. It simply indexes which, among the several irreducible representations of the group, we are considering. Both $\kappa$ and $k$ span the same *small* number of values - called the dimension of the representation. Irrep labels catalogue the invariants perceivable under the transformations of the group. When the observer performs her altered scrutinies she notices what remains the same among her observations. If she labels the object by these sameness regimes she is effectively using irrep labels.

The crucial property of *organization by invariant behaviour* is contained in this fundamental equation of group theory

$$G_g \left| \gamma, k... \right\rangle = \sum_\kappa \left| \gamma, \kappa... \right\rangle \left\langle \kappa \left| \Gamma_g^\gamma \right| k \right\rangle \qquad (5)$$

It says that there are states, labelled by $\gamma$, which have the special property that under *all altered scrutinies* only states labelled by the same $\gamma$ appear. (The dots in $\left| \gamma, k... \right\rangle$ allow for other state markers that are commensurate with its $\gamma$ property. They don't affect its $\gamma$-ness.) When the observer thinks in terms of these states she is perceiving a symmetry in the system - perceiving the sameness features present under altered scrutiny and labelling them by $\gamma$.

## 12. Identity resides in labels of significance

Here is a primitive example. For a square there are eight altered scrutinies that preserve apparent sameness. They make up the group D$_4$, the symmetry group of a square. The essential feature which enables us to attach the label 'square' to something is that it transforms as the first irreducible representation of the group D$_4$. In this resides its squareness: that it admits of the label γ(D$_4$) = 1 (A$_1$ in crystallographic notation). The irreducible representation label, γ(D$_4$)=A$_1$, is an alternative name for squareness.

Thus the symmetry results in an identifying name. What group theory does is to produce *names* for *behaviour*; behaviour under group transformations. The names are the irreducible representation labels. These labels have the significance that they name patterns in the appearance changes that arise from altered scrutiny.

Not all labels are 'labels of significance' and hence of identity. Labels of enumeration do not identify, except primitively - for mere presence. The collection of states $|n\rangle$ are labelled by enumeration. The collection of states $|\gamma, k...\rangle$ are labelled by significance.

Just as any state at all is a superposition of the whole gamut of basis states, $|n\rangle$, so too is $|\gamma, k...\rangle$ such a superposition. The $\gamma$-basis is another way of describing the system. Hence there exists a transformation matrix, $U$, with elements $\langle n|\gamma, k...\rangle$ by which symmetry basis states, $|\gamma, k..\rangle$ are derived from enumerative basis states, $|n\rangle$. Group theory provides us a prescription for getting such $U$ matrices.

Structurally $U$ can be viewed as an altered scrutiny operator. Any state $|n\rangle$, under altered scrutiny $U$, appears as $U|n\rangle = |\gamma, k...\rangle$. The observer who alters her scrutiny in this special way perceives the identifying symmetry in the system. Thus the act of assigning a $\gamma$-label is associated with an intellectual transformation. To assign a label of significance amounts to a change of basis from enumerative labeling to significance labeling. To grant a label of significance is as much attached to understanding as to identification. Granting such a label is the signature of understanding.

## 13. The full symmetry of the state is a sub-symmetry of the rule

A system is governed by a rule. And this rule has its symmetry. The group of all observer transformations that accommodate the rule is the *group of the rule* of the system. In physics the rule is expressed via a Lagrangian, a Hamiltonian or a variational principle. The states of the system derive from the rule. But the symmetry of any state of the system is not necessarily that of the rule!

An extension of the square example illustrates it. Suppose, instead of a square, the system is a rectangle - a stretched square. The system has states; rectangles - with long sides horizontal or long sides vertical. Each of these states, when examined under altered scrutiny, still obeys the rule. A rectangle results - albeit in different orientations. A $D_4$ altered scrutiny produces a different state but does not contravene the rule of the system. Executing all eight altered scrutinies of $D_4$ the observer finds that what she sees is characterizable by the symmetry label $\gamma(D_4) = 3$ ($B_1$ in spectroscopic notation). Put formally, the state with this label transforms as the third irreducible representation of $D_4$.

But the particular object she is examining has $D_2$ as its group of 'symmetry operations'. For only four of the eight altered scrutinies of $D_4$ does it look the same: a 180° rotation about the principal axis and about each of the two perpendicular axes plus the do-nothing altered scrutiny. A state has a lower symmetry than the rule governing the system. Such lower symmetry states are said to exhibit *broken symmetry*.

For any group there is always one particular value of γ that expresses its full symmetry: the value of $\gamma$ for which $\Gamma_g^\gamma$ is one-dimensional and equal to 1 for all g. Thus in equation (5) the full-symmetry $\gamma$ marks the case $\kappa = k = 1$ and $\langle \kappa | \Gamma_g^\gamma | k \rangle = 1$. Full symmetry means that for any g at all

$$G_g | \gamma = 1... \rangle = | \gamma = 1... \rangle \qquad (6)$$

Traditionally the full-symmetry γ-value is taken as zero for continuous groups and unity for groups with countable elements. All other values of γ are sub-symmetry ones. The rectangle example illustrates the important fact that to each sub-symmetry value of $\gamma$ is associated the full-symmetry of a subgroup. To $\gamma(D_4) = 3$ (or $B_1$) is associated $\gamma(D_2) = 1$ (or $A_1$).

Planetary motion provides the best-known example involving dynamical symmetries. Neither the coulomb nor the gravitational force holding a mass in orbit have an orientation dependence. The Hamiltonian is invariant under all the transformations of SO(3). Thus the *rule* of the system has SO(3) symmetry. But the orbits of this system do not. They lie in a plane having at best O(2) symmetry - a subgroup of SO(3).

In the Bohr atom the irrep label is the angular momentum quantum number: $\gamma = \ell$. A state of the system carries this irrep label. To each $\ell > 0$ corresponds a subgroup of SO(3) which characterizes the full symmetry of the state. So the full symmetry of the state is that of a sub-symmetry of the rule.

A cornucopia of examples from many disciplines - Couette flow, vortices, Sequoia tree bark, corn circles, crystal growth, Chladni patterns - all exemplifying this principle are charmingly and informatively presented without equations in *Fearful Symmetry* (Stewart and Golubitsky, 1992).

Reference is made there to Curie's Principle. It is summarized as "symmetric causes produce equally symmetric effects" (Stewart and Golubitsky, 1992, p. 8). As abundantly documented in the book this is not true. What is true is that the symmetry of the state (the effect) is a subgroup of the symmetry of the rule (the cause). Symmetry is not so much about cause and effect as it is about rules and states of a system. That Curie's Principle is incorrect has been widely recognized. (See Roche, 1987, p. 20, Stewart and Golubitsky, 1992 and Radicati, 1987, pp. 197-206)

## 14. Depth of inspection is a larger group

The significance of the previous section - the mathematical connection between rule symmetries and state symmetries - is this. Our ideas about what we know (the state symmetry) can often be fit into a larger concept (the rule symmetry). The identification process depends upon the depth of inspection - the range of altered perspectives pursued by the observer. In inspecting the system the choice of altered scrutinies is the observer's. The label she assigns to a system heralds her range of scrutinies. For example any one rectangle has $D_2$ symmetry (four scrutinies). But the rule has $D_4$ (eight scrutinies). Identity evolves as the range of scrutinies expands. This feature introduces a suggestion of subjectivity. What one observer may perceive is not what another would because each observer has her own range of altered scrutinies.

The parallel mathematics is this: The group of all transformations to which the rule is invariant characterizes the system. The states of the system carry the irrep labels of this group. But these are generally sub-symmetry $\gamma$-values. They correspond to full symmetries only of subgroups of the rule group. An observer may not, in fact, examine all the scrutinies the rule makes available. He perceives, with his limited imagination for scrutinies, only invariance to a subgroup. He tags the system with the irreps of this subgroup. These mark its identity to this observer. Until he widens his perspective he may not perceive the system's full identity. The process of identification is a matter of acheiving perspective. It parallels the mathematics of discovering a larger group into which the facts fit.

An implication is that *inference* - how we evaluate perception - is a matter of this insight: that observations, at first thought to constitute a group, are properly a subgroup among a broader constellation of observations. In the planetary system, cited earlier, each circular orbit observed has SO(2) symmetry. But taken together these observations issue from an SO(3) rule. Labelling the orbits by SO(3) exhibits this deeper perception.

## 15. Understanding: perceiving accidental as normal

In physics this process relates to the phenomenon of 'accidental degeneracy'. The term refers to the situation where two or more states with *different irrep label values* have the same energy. In this event the group yielding these irrep labels does not account for the degeneracy. It doesn't force the degeneracy to exist. A multidimensional irreducible representation, by its nature, generates multiplets of energy-degenerate states. Such degeneracy is called 'normal'. It is 'normal' because it is forced by the rule's symmetry. States must exhibit a degeneracy equal to the dimension of their irreducible representation. If the degeneracy embraces states with *different irrep labels* then the degeneracy is not forced by the symmetry. It is called accidental degeneracy meaning unexplained degeneracy.

Explanation is sought in 'hidden symmetry'. This amounts to the conclusion that our inspection is too circumscribed. We are not discerning features of the system that necessitate the degeneracy we actually measure. A grander group of scrutiny alternatives is needed. The search for hidden symmetry[11] consists in seeking a larger group of altered scrutinies - a group of higher order or of more parameters than the original. Of these, the ones we already use are a subgroup. The goal is to discover a larger group for which the observed degeneracy is normal; to reveal the single irrep label that subsumes all of the degenerate states. Understanding consists in perceiving accidental degeneracy as normal.[12]

The prototypical example of this process concerns the model spinless hydrogen atom; bound state energies = -13.6 eV/n$^2$. [13] All of its levels above the ground state have accidental degeneracy. They depend only upon the principal quantum number, n, and not on the angular momentum quantum number, $\ell$. But it is precisely $\ell$ that names the symmetry properties of the states. The $\ell = \gamma$ are the irrep labels of the group of all the altered scrutinies of rotation in real three-dimensional space, SO(3).

The situation is exemplified in the first excited state, n=2. It is four-fold degenerate. The $l=0$ state and the three $l=1$ states have the same energy. A normal degeneracy would have the three $l=1$ states with an energy different from that of the $l=0$ state. That they are not different is the accidental degeneracy.

The resolution lay in expanding scrutinies beyond the geometrical to include the dynamical - the variables of motion. With this expansion in perspective it can be shown that O(4), like SO(3), is a symmetry group of the model atom Hamiltonian where the higher O(4) symmetry contains SO(3) as a subgroup. Whereas SO(3) is a three-parameter group, O(4) has six parameters. There are many more altered scrutinies associated with O(4) - difficult to physically execute but implementable conceptually. The generators for SO(3) are the angular momentum operators. To each of these correspond angles which are the group parameters - geometric ones. For O(4) there are three more generators. They express the constancy of the Runge-Lenz vector. The constancy of this vector corresponds to the classical finding that the eccentricity of the orbit is unvarying. The orbit doesn't precess or distort during the course of motion. These are sameness features under all scrutinies. For O(4), n itself, is an irrep label with degeneracy $n^2$. This is precisely the degeneracy of the system so under O(4) the degeneracy is normal. Under SO(3) it is accidental.

Understanding consisted in a conceptual metamorphosis from perceiving an accidental degeneracy to perceiving a normal one. The process required a broader set of scrutinies. One might liken it to the perception of nuance as a measure of understanding.

## 16. Conservation laws yield symmetry labels

The relation between conservation laws and symmetry has a long history (Kastrup, 1987). It culminated in Noether's Theorem which is nicely synopsized by Mills (1989, p. 494) as "for every symmetry of nature there is a corresponding conservation law and for every conservation law there is a symmetry of nature".

This fundamental theorem can be seen as a consequence of the notion that symmetry is identity. 'Conserved' is what remains the same under altered scrutinies. To identify is to recognize something that is conserved. Whatever label identifies something is what marks the thing that is conserved. But the assignment of an identifying label is exactly what symmetry produces. So, indeed, what is conserved and what has symmetry are the same thing; they are identified via an irrep label.

## 17. Irrep labels are names for behaviour

The elements of the periodic table are named by what is conserved in all the chemical reactions among them: the atomic number, Z. Atomic number is a label for the number of electrons in the system. That this is conserved means that you can label with it. The quantum numbers by which we identify the states of an atom of one of these elements are its irrep labels. They arise from generators which catalogue the irreducible representations of a Lie group. The same applies to the elementary particles of particle physics. They, too, are characterized by conserved quantities which astute physicists associate with abstract groups whose algebra fits observations. The particle data tables of high energy physics (Groom, D. E. et al., 2000) list the association between the names of particles and their quantum numbers. But each quantum number is a label or index for the irrep of a group. Many of them come from Lie group generators such as angular momentum and energy. Others relate to finite groups like parity. Particles are named by their irrep labels in these tables.

Electrical charge, being a conserved quantity, is the irrep label for sameness of the electric and magnetic fields under changes in their scalar and vector potentials. The altered scrutinies are the gauge transformations that preserve the fields.

The designation 'middle-A' for a musical note is another name for the irrep label $\gamma = 440 Hz$ of the group of translations in time. It characterizes a signal which, under the group of all altered scrutinies in time by integer multiples of 2.3 milliseconds, appears (sounds) the same to the observer. By the same token the color red is a surrogate for the irrep label $\gamma = 4.3 \times 10^{14} Hz$. In fact a Fourier transform is simply a signal represented in the symmetry basis states of the translation group.

That 'edgeness' has a symmetry name was used by Lenz (1989) in a cleverly practical way. He found that edgeness can be characterized by the group U(1). Any particular edge has a $\gamma$-value where $\gamma$ is the eigenvalue corresponding to the Lie group generator for U(1) and thus an irrep index for the group. He used this information in programming a computer to determine edgeness in any 2-D region of a picture. He calls his work "feature extraction". It identifies a feature.

## 18. Is symmetry identity?

The preceding sections presented the case for the structural parallel between symmetry and identity. Portraying group theory as the formal expression of our elemental experience of nature constitutes the case that the effectiveness of mathematics is reasonable. To validate the symmetry-identity connection examples abound where an identifying name marks an irrep label. Nevertheless that 'symmetry *is* identity' has not been proved. Many identifications have not been shown to fit the scheme. Symmetry is sufficient for identification but may not be necessary.

What then - beyond sameness under altered scrutiny - would be necessary? I cannot say. But if symmetry were necessary then there must be an irrep for "shoeness". And in fact irreps for all notions already identified by people. Until we can perceive the irrep label that characterizes how we perceive shoeness in the objects we put on our feet we cannot be sure that symmetry is, indeed, identity.

## Notes

[1] Discussions of this point are in d'Espagnat (1976, p. 14) and in Rosen (1995, p. 66).

[2] Cohen-Tannoudji, Diu, and Laloë (1977) is a good standard text on the quantum mechanics needed here.

[3] Appendix A of Jones (1994) gives a unified overview of quantum mechanics expressed solely in Dirac notation. That Dirac notation encodes physical meaning is a central theme in Chester (1987)

[4] Note that the word 'invariant' in 'invariant space' is not to be confused with invariance of a state. A state in an invariant space is not generally invariant to a group transformation.

[5] An example is given on p. 10 of Chaichian and Hagedorn (1998).

[6] Considerable attention is devoted to this by Chaichian and Hagedorn (1998) on p.17 and then again in Chapter 3.

[7] "... symmmetry operations involve a movement of the body.", p. 3 of Burns (1977)

[8] Induced transformation is discussed at length in Section 3.7 of Elliott and Dawber (1979).

[9] Group theory books with the word symmetry in their titles pay attention to definition. Rosen (1995, p. 157) defines symmetry as "immunity to a possible change". Chaichian and Hagedorn (1995, p. 17) give a definition but in technical language and too long to reproduce here. Elliott and Dawber (1979) quote the dictionary but do not relate this definition to the mathematics. Weyl (1952) treats us to poetry and history on the word and then uses it eloquently as if it were defined. He finally accepts the word congruence for its essential property. An "...automorphism of a generalized function space ..." says Redhead (1975).

[10] A pithy but complete summary of the structure of quantum mechanics is given by Weinberg (1995, pp. 49-50).

[11] McIntosh (1971, pp.80-84) gives an early history of this pursuit.

[12] This pursuit was not always considered worthy. Wigner (1959, p.120) dismissed it with, "It will be assumed that accidental degeneracy is a very uncommon situation and that it occurs ... only exceptionally."

[13] It is called the non-relativistic Kepler problem by Barut and Raczka (1986, p. 381). Very fine expositions on this seminal excercise are also given in Jauch and Hill (1940), Joshi (1985, p.177-182) and Jones (1994, p.124-127).